\documentclass[%
 reprint,
superscriptaddress,
nofootinbib,
nobibnotes,
 amsmath,amssymb,
 aps, prd
]{revtex4-2}

\usepackage{graphicx}% Include figure files
\usepackage{dcolumn}% Align table columns on decimal point
\usepackage{bm}% bold math
\usepackage{hyperref}% add hypertext capabilities
\usepackage{aas_macros}% Include figure files
\usepackage{orcidlink}

\newcommand{\UCB}{\affiliation{Department of Physics, University of California, Berkeley, Berkeley, CA 94720, USA}}
\newcommand{\UNH}{\affiliation{Department of Physics \& Astronomy, University of New Hampshire, 9 Library Way, Durham NH 03824, USA}}
\newcommand{\WSU}{\affiliation{Department of Physics \& Astronomy, Washington State University, Pullman, Washington 99164, USA}}
\newcommand{\Caltech}{\affiliation{Theoretical Astrophysics, Walter Burke Institute for Theoretical Physics, California Institute of Technology, Pasadena, California 91125, USA}}

\newcommand{\Cornell}{\affiliation{Cornell Center for Astrophysics and Planetary Science, Cornell University, Ithaca, New York 14853, USA}}
\newcommand{\AEI}{\affiliation{Max Planck Institute for Gravitational Physics (Albert Einstein Institute), Am M\"uhlenberg 1, D-14476 Potsdam, Germany}}

\graphicspath{{./}{figures/}}

\begin{document}

\preprint{APS/123-QED}

\title{High angular momentum hot differentially rotating equilibrium star evolutions in conformally flat spacetime}% Force line breaks with \\

\author{Patrick Chi-Kit \surname{Cheong}~\orcidlink{0000-0003-1449-3363}}
\email{patrick.cheong@berkeley.edu}
\altaffiliation[]{N3AS Postdoctoral fellow}%Lines break automatically or can be forced with \\
\UNH
\UCB

\author{Nishad Muhammed~\orcidlink{0000-0001-8574-0523}} \WSU
\author{Pavan Chawhan~\orcidlink{0000-0002-3694-7138}} \WSU
\author{Matthew D. Duez~\orcidlink{0000-0002-0050-1783}} \WSU
\author{Francois Foucart~\orcidlink{0000-0003-4617-4738}} \UNH
\author{Lawrence E.~Kidder~\orcidlink{0000-0001-5392-7342}} \Cornell
\author{Harald P. Pfeiffer~\orcidlink{0000-0001-9288-519X}} \AEI
\author{Mark A. Scheel~\orcidlink{0000-0001-6656-9134}} \Caltech

% Patrick Chi-Kit Cheong and Nishad Muhammed and Pavan Chawhan   and Matthew D. Duez and Francois Foucart

\date{\today}% It is always \today, today,

\begin{abstract}
The conformal flatness approximation to the Einstein equations has been successfully used in many astrophysical applications such as initial data constructions and dynamical simulations.
% Although it has been shown that full general relativistic strongly differentially rotating equilibrium models deviate by at most a few percents from their conformally flat counterparts, whether those solutions share the same dynamical stabilities has not been fully addressed.
Although it has been shown that full general relativistic strongly differentially rotating equilibrium models deviate by at most a few percent from their conformally flat counterparts, whether those conformally flat solutions remain stable has not been fully addressed.
To further understand the limitations of the conformal flatness approximation, in this work, we construct spatially-conformally-flat hot hypermassive neutron stars with post-merger-like rotation laws, and perform conformally flat evolutions and analysis over dynamical timescales.
% We find that the stellar profiles of quasi-toroidal models with high angular momentum for $J \gtrsim 9 ~G M_{\odot}^2 / c$ can change significantly over dynamical timescales.
We find that enforcing conformally-flat spacetime could change the equilibrium of quasi-toroidal models with high angular momentum for $J \gtrsim 9 ~G M_{\odot}^2 / c$ compared to fully general relativistic cases.
In contrast, all the quasi-spherical models considered in this work remain stable even with high angular momentum $J=9~G M_{\odot}^2 / c$.
Our investigation suggests that the quasi-spherical models are suitable initial data for long-lived hypermassive neutron star modeling in conformally flat spacetime.
% Our investigation suggest that the dynamical stabilities of high momentum quasi-equilibrium models can be non-trivial in conformally flat spacetime.
% Nevertheless, the post-merger like quasi-spherical models are ideal initial conditions.
% \begin{description}
% Secondary publications and information retrieval purposes.
% You may use the \texttt{description} environment to structure your abstract;
% use the optional argument of the \verb+\item+ command to give the category of each item. 
% \end{description}
\end{abstract}

\maketitle

%\tableofcontents

% \texttt{Gmunu}~\cite{2020CQGra..37n5015C, 2021MNRAS.508.2279C, 2022ApJS..261...22C, 2023ApJS..267...38C},~\cite{2023arXiv230903526H}

\section{\label{sec:intro}Introduction}
The detection of a binary neutron star merger on 17 August 2017 has laid a milestone in multi-messenger astronomy.
This event was observed by the coincident detections of gravitational waves GW170817~\cite{2017PhRvL.119p1101A}, the short gamma-ray burst GRB170817A~\cite{2017ApJ...848L..13A}, and in other spectral bands~\cite{2017ApJ...848L..12A}.
Even though this groundbreaking multimessenger detection has confirmed our basic understanding of neutron star mergers~\cite{2017arXiv171005931M, 2018ASSL..457.....R}, details of the post-merger evolution are poorly understood.
A hypermassive neutron star, which is expected to be hot and supported by strong differential rotation, is one of the possible outcomes of binary neutron star merger. 
Studying hypermassive neutron star helps us to further understand the nature of the central engine of the relativistic jets~\cite{2014ARA&A..52...43B,2017ApJ...848L..13A} and kilonova transients~\cite{1998ApJ...507L..59L, 2010MNRAS.406.2650M, 2016AdAst2016E...8T, 2019LRR....23....1M}.

Detailed investigations of the post-merger phase over dynamical and secular timescales are extremely challenging yet significant.
Not only does one need to solve Einstein field equations and general-relativistic magneto-hydrodynamics self-consistently~\cite{2018ASSL..457.....R}, neutrino microphysics is also required~\cite{2023LRCA....9....1F}.
Moreover, to better understand the post-merger observational signatures, seconds-long simulations are required.
Hence, any simplification of the simulations is highly desirable.

The spatially conformally flat spacetime approximation~\cite{wilson_mathews_2003, 2008IJMPD..17..265I, xCFC} has shown to be useful for modeling neutron star mergers.
Binary neutron star merger simulations based on the conformally flat approximation have been successfully carried out (e.g.~\cite{2002PhRvD..65j3005O, 2012PhRvD..86f3001B, 2013ApJ...773...78B, 2014PhRvD..90b3002B, 2021PhRvD.103l3004B, 2024MNRAS.528.1906L}, see also the applications in the context of core-collapse supernovae~\cite{2002A&A...393..523D, 2004ApJ...615..866S, 2015MNRAS.453..287M} and isolated neutron stars~\cite{2006MNRAS.368.1609D, XECHO, 2021ApJ...915..108N, 2022CmPhy...5..334L, 2023arXiv230316820Y, 2024arXiv240113993Y, 2011A&A...528A.101B,2014MNRAS.439.3541P,2015MNRAS.447.2821P,2017MNRAS.470.2469P,2020A&A...640A..44S}).
In addition, it has been shown in fully general relativistic simulations that long-lived neutron star merger remnants are qualitatively axisymmetric, and the corresponding spacetime is nearly conformally flat~\cite{2017ApJ...846..114F, 2023arXiv231211358H}. 
Mapping such post-merger profiles by assuming conformally flat conditions onto other evolution codes that impose different symmetries or with different input physics has been done recently~\cite{2017ApJ...846..114F, 2023arXiv231211358H}.
Specifically, the multigrid based conformally-flat spacetime solver of \texttt{Gmunu}~\cite{2020CQGra..37n5015C, 2021MNRAS.508.2279C} has been demonstrated to be very effective for the studies of long-lived post-merger neutron star merger remnants over secular timescales~\cite{2023arXiv231211358H}.
% fixme: maybe add a section about why construct equilibrium and not BNS

% fixme: Maybe more clearly state why conformally flat simulations are useful / why we need to know how reliable they are.
Understanding the limitation of the conformally flat spacetime approximation in the context of hypermassive neutron stars is critical.
Despite the success in neutron star modeling (e.g.~\cite{2002PhRvD..65j3005O, 2012PhRvD..86f3001B, 2013ApJ...773...78B, 2014PhRvD..90b3002B, 2021PhRvD.103l3004B, 2024MNRAS.528.1906L}), the conformally flat condition is ultimately an approximation.
This approximation is no longer valid in the case of a Kerr black hole~\cite{2000PhRvD..61l4011G, 2004PhRvL..92d1101K}, and it may also fail with systems that have extreme rotation or high angular momentum.
Studies have shown that the local and integrated quantities are at most a few percent difference between fully general relativistic and conformally flat differentially rotating equilibrium models~\cite{1996PhRvD..53.5533C, 2014GReGr..46.1800I, 2021MNRAS.503..850I, 2022MNRAS.510.2948I}.
%~\cite{1999PhRvD..60b7501K} for disk
However, whether the full and conformally flat solutions share the same properties of dynamical and secular stabilities is not clearly addressed.
Recently, the conformally flat approximated dynamical evolutions of quasi-toroidal models with the $J$-constant rotation law~\cite{1988ApJ...325..722F} with polytropic equation of state has been carried out~\cite{2023arXiv230206007S}.
Nevertheless, the rotation law considered in~\cite{2023arXiv230206007S} is very different from that of post-merger remnants.
It is still unclear whether post-merger like hypermassive neutron stars can be accurately modelled under the conformal flatness approximation.

In this work, we investigate the limitations of the conformally-flat approximation in high angular momentum post-merger-like hypermassive neutron star modeling.
% In this work, we investigate the limitations of using axisymmetric differentially rotating quasi-equilibrium models to be initial conditions for conformally flat hypermassive neutron star modeling.
% To do this, we challenge the conformally flat approximation simulations with high angular momentum merger-like neutron stars.
In particular, we construct spatially-conformally-flat post-merger-like hot hypermassive neutron stars, and perform evolutions and analysis over dynamical timescales.
We find that the stellar profiles of conformally-flat quasi-toroidal models with high angular momentum for $J \gtrsim 9 ~G M_{\odot}^2 / c$ can be distorted noticeably over dynamical timescales even in fully general relativistic evolutions.
However, the fully general relativistic variant of such stars remain stable in fully general relativistic evolutions within the time we simulated.
This implies that conformally-flat approximation either makes such high angular {momentum} star not an equilibrium or makes it an unstable equilibrium.
%fixme: Since this phenomenon is the main finding of this paper, we should make some effort to determine which of the two possibilities is the case.
On the other hand, all the quasi-spherical models considered in this work remain stable even with high angular momentum $J=9~G M_{\odot}^2 / c$.
Our study suggests that the quasi-spherical models are better choice to be used as hypermassive neutron star modeling because of their rotation properties and stabilities.
% Our study suggests that the dynamical stability of high momentum quasi-equilibrium models can be non-trivial in conformally flat spacetime.

The paper is organised as follows.
In section~\ref{sec:methods} we outline the methods we used in this work.
The results are presented in section~\ref{sec:results}.
This paper ends with a discussion in section~\ref{sec:discussion}.
Unless explicitly stated, we use the units in which the speed of light $c$, gravitational constant $G$, solar mass $\rm{M_{\odot}}$ are all equal to one ($c=G={\rm M_{\odot}}= 1$).
% Greek indices, running from 0 to 3, are used for 4-quantities while the Roman indices, running from 1 to 3, are used for 3-quantities.

\section{\label{sec:methods}Methods}
\subsection{\label{sec:id}Initial conditions}
% Conformally flat, axisymmetric, differentially rotating hot neutron stars quasi-equilibriums are constructed by utilising \texttt{RotNS} code~\cite{1994ApJ...422..227C}, and serve as our initial data.
Conformally flat, axisymmetric, differentially rotating hot neutron stars in quasi-equilibrium are constructed using the \texttt{RotNS} code~\cite{1994ApJ...422..227C}, and serve as our initial data.
\texttt{RotNS}~\cite{1994ApJ...422..227C} was used to construct equilibrium sequences of rotating polytropes in general relativity. 
The code has been recently updated to support tabulated equations of state and the 4-parameter rotation law of {Ury{\={u}}} \emph{et~al.}~\cite{2019PhRvD.100l3019U}.
For the implementation details, we refer readers to~\cite{2024arXiv240305642M}.
Below, we will only highlight the key setup of the initial data construction.

Although \texttt{RotNS}~\cite{1994ApJ...422..227C} is a fully general relativistic code, the spatially-conformally-flat conditions can easily be enforced by imposing an additional condition of the metric potentials, as shown in~\cite{1996PhRvD..53.5533C, 2014GReGr..46.1800I, 2021MNRAS.503..850I, 2022MNRAS.510.2948I}.
Here we adopt the same modification in \texttt{RotNS} to construct conformally flat initial data.
%fixme: I think these modifications should be reproduced here since they are central to the topic of the paper.
Unless explicitly stated, all the initial data are constructed in conformally-flat spacetime.

To construct merger-like hypermassive neutron star profiles, we adopt the 4-parameter rotation law of {Ury{\={u}}} \emph{et~al.}~\cite{2019PhRvD.100l3019U}.
In particular, we implement the following rotation law,
\begin{equation}
\label{eq:uryu-general}
    \Omega\left(j ; \Omega_c\right)=\Omega_c \frac{1+\left[j /\left(B^2 \Omega_c\right)\right]^p}{1+\left[j /\left(A^2 \Omega_c\right)\right]^{q+p}},
\end{equation}
where $j$ is the specific angular momentum, $\Omega_{\rm c}$ is the central angular velocity of the star, while $A$, $B$, $q$, and $p$ are parameters.
In this work, we choose $p=1$ and $q=3$.
Note that this rotation profile is non-monotonic, with the maximum angular velocity $\Omega_{\max}$ between the centre and surface.
The characteristic of the models can be controlled by specifying parameters $A$ and $B$.
Alternatively, parameters $A$ and $B$ can be obtained by fixing angular velocity ratios ${\Omega_{\max}}/{\Omega_{\rm c}}$ and ${\Omega_{\rm eq}}/{\Omega_{\rm c}}$~\cite{2019PhRvD.100l3019U, 2021MNRAS.503..850I, 2022MNRAS.510.2948I}, where $\Omega_{\rm eq}$ is the equatorial angular velocity of the star. 
Different choices of the angular velocity ratios can result in either quasi-toroidal or quasi-spherical models.

\begin{figure}
	\centering
	\includegraphics[width=\columnwidth, angle=0]{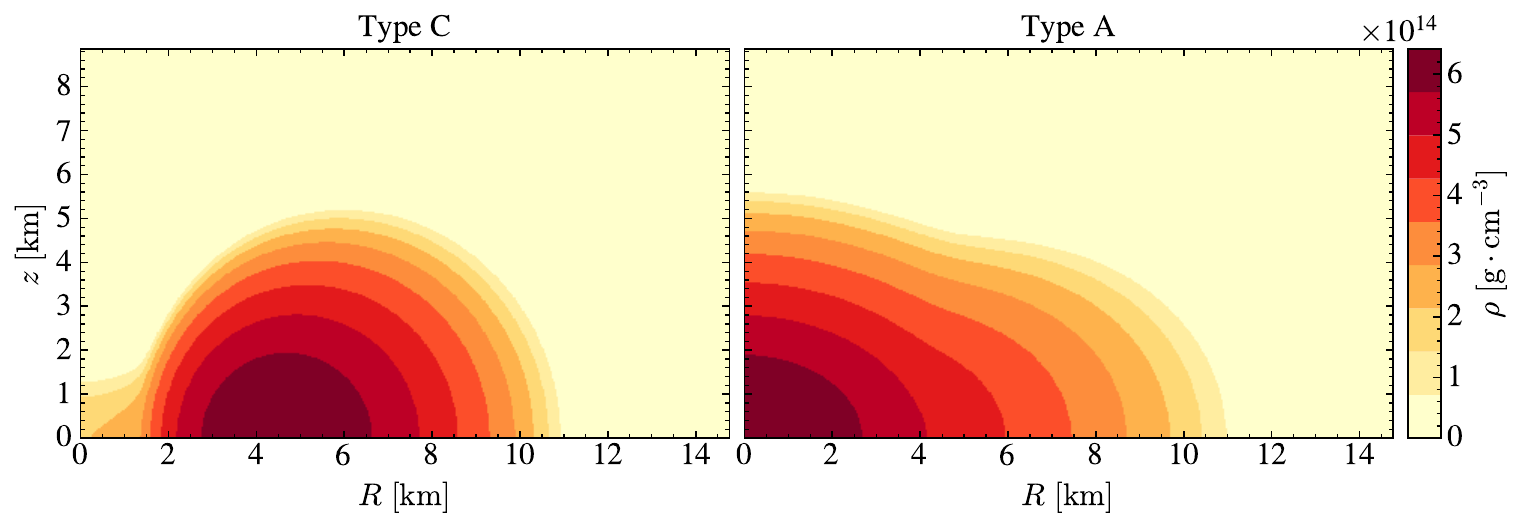}
	\caption{
		Two-dimensional rest mass density profiles $\rho$ in the units of $\rm{[g \cdot cm^{-3}]}$ of the rotating neutron star models.
		\emph{Left panel}: Quasi-toroidal (type C) model with $\left\{{\Omega_{\max}}/{\Omega_{\rm c}}=2,\;{\Omega_{\rm eq}}/{\Omega_{\rm c}}=0.5\right\}$.
		This star has the angular momentum $J=11~G M_{\odot}^2 / c $ with the maximum energy density $\epsilon_{\max} = 7.073\times10^{14}~\rm{g \cdot cm^{-3}}$.
		\emph{Right panel}: Quasi-spherical (type A) model with $\left\{{\Omega_{\max}}/{\Omega_{\rm c}}=1.6,\;{\Omega_{\rm eq}}/{\Omega_{\rm c}}=1\right\}$.
		This star has the angular momentum $J=9~G M_{\odot}^2 / c $ with the maximum energy density $\epsilon_{\max} = 9.411\times10^{14}~\rm{g \cdot cm^{-3}}$.
		}
	\label{fig:rho_ns}	
\end{figure}

In this work, we consider two sets of the ratios $\left\{{\Omega_{\max}}/{\Omega_{\rm c}},\;{\Omega_{\rm eq}}/{\Omega_{\rm c}}\right\}$, namely: 
(i) $\left\{2,0.5\right\}$ which results in quasi-toroidal models, and
(ii) $\left\{1.6,1\right\}$ which results in quasi-spherical models (see figure~\ref{fig:rho_ns} for examples of the rest mass density profiles of both types of stars).
They can be classified as type C and type A solutions according to~\cite{2009MNRAS.396.2359A}.
% Note that the former choice of the ratios is widely used but has no astrophysical motivations.
% On the contrary, the latter set of the ratios are chosen to match the results of the numerical relativity simulations of binary neutron star mergers (e.g. )~\cite{2021MNRAS.503..850I, 2022MNRAS.510.2948I}.
Note that the latter set of the ratios are chosen to match the results of the numerical relativity simulations of binary neutron star mergers (e.g.~\cite{2017PhRvD..96d3004H, 2020PhRvD.101f4052D}, see~\cite{2021MNRAS.503..850I, 2022MNRAS.510.2948I}).
% Therefore, the quasi-spherical type models are by construction very similar to the binary neutron star merger remnants.
Therefore, the quasi-spherical type models are by construction more ``post-merger-like'' compared to quasi-toroidal models.

All the equilibrium models in this work are constructed with equation of state DD2~\cite{2010NuPhA.837..210H} with a constant entropy per baryon $s = 1 ~ k_{\rm{B}} / \text{baryon}$ and in neutrinoless $\beta$-equilibrium.
The temperature is roughly 30~MeV at the centre of the star.
Note that, such choice of entropy profile and the resulting temperature profile does not match the numerical relativity simulations, where the temperature at the centre is expected to be lower than the surface.
Nevertheless, we consider only the constant entropy profile for simplicity.
The investigations of different choice of entropy profiles will be left as future work.
% fixme: define compacted radius, and states that this is in RNS
The resolution of the compacted radius and angular grid (see their definitions in~\cite{1994ApJ...422..227C}) in \texttt{RotNS} is $600 \times 600$.

\subsection{\label{sec:sim}Simulation setup}
We employ the general relativistic magnetohydrodynamics code \texttt{Gmunu}~\cite{2020CQGra..37n5015C, 2021MNRAS.508.2279C, 2022ApJS..261...22C, 2023ApJS..267...38C, 2023arXiv230903526H} to evolve the neutron star models in dynamical conformally-flat spacetime. 
All the simulations here are axisymmetric (i.e. 2-dimensional) in cylindrical coordinates $(R, z)$, where the computational domain covers $0 \leq R \leq 120$ and $0 \leq z \leq 120$, with the resolution {$n_R \times n_z = 128 \times 128$} and allowing 6 AMR levels.
The finest grid size at the centre of the star is $\Delta R = \Delta z \approx 43.27 ~\rm{m}$.
% The refinement setting is the same as in our previous work~\cite{2021MNRAS.508.2279C}.
% In particular, we defined a relativistic gravitational potential $\Phi \equiv 1 - \alpha$.
% For any $\Phi$ larger than the maximum potential $\Phi_{\text{max}}$ (which is set as 0.2 in this work), the block is set to be finest.
% While for the second-finest level, the same check is performed with a new maximum potential which is half of the previous one, so on and so forth.
The refinement is fixed after the initialisation since we do not expect the stars expand significantly.

Our simulations adopt Harten, Lax and van Leer (HLL) approximated Riemann solver~\cite{harten1983upstream}, 3rd-order reconstruction method PPM~\cite{1984JCoPh..54..174C} and 3rd-order accurate SSPRK3 time integrator~\cite{1988JCoPh..77..439S}. 
Finite temperature equation of state DD2~\cite{2010NuPhA.837..210H} is used for the evolutions.
Although neutrinos are not included, the electron fraction $Y_e$ is evolved in these simulations.

{
The rest-mass density of the atmosphere $\rho_{\rm atmo}$ is set to be $10^{-10} \rho_{\max}\left(t=0\right)$.
For anywhere that the matter has rest-mass density lower than $\rho_{\rm atmo}$, we reset the rest-mass density of those regions to be $0.2 \rho_{\rm atmo}$, and zero the velocities (i.e. $v^i = 0$).
As a result, the angular velocity $\Omega \equiv \alpha v^{\phi} - \beta^{\phi}$ has a suddent drop at the neutron star surface at $t=0$ (see e.g. figure~\ref{fig:uryu_2.0_0.5_var_e}).
These areas will be filled with low density gas that rotates with similar angular velocity as soon as the simulation started, and do not affect the dynamics of the neutron star because of the ultra-low rest-mass density.
}

To initialise the simulations, we map the conservative variables of the stars into \texttt{Gmunu}, and solve the metric again with the multigrid metric solver~\cite{2020CQGra..37n5015C, 2021MNRAS.508.2279C}.
This approach can also be applied on fully general relativistic profiles, which was used in~\cite{2017ApJ...846..114F, 2023arXiv231211358H}.

\section{\label{sec:results}Results}
\subsection{\label{sec:j_seq}Sequences of equilibrium models}
\begin{figure*}
	\centering
	\includegraphics[width=\textwidth, angle=0]{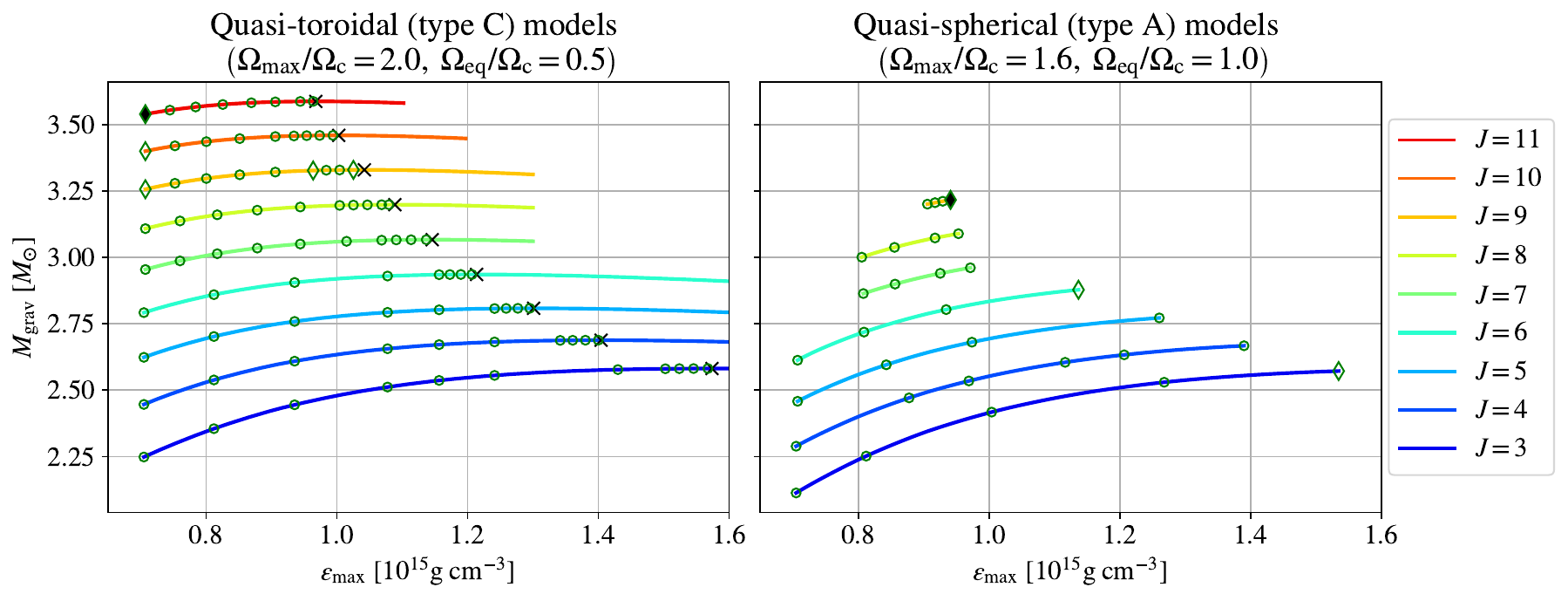}
	\caption{
		Gravitational mass $M_{\rm{grav}}$ versus maximum energy density $\epsilon_{\max}$ for various constant angular momentum sequences.
		Circles and diamonds mark the dynamically evolved models without introducing perturbations.
		The diamonds refer to the models presented in detail in figure~\ref{fig:uryu_2.0_0.5_var_e}, \ref{fig:uryu_2.0_0.5_var_j} and \ref{fig:uryu_1.6_1.0_var_j}.
		The black diamonds are the two cases presented in figure~\ref{fig:rho_ns}.
		None of the simulations presented here collapse as black holes.
		\emph{Left panel}: Quasi-toroidal (type C) model with $\left\{{\Omega_{\max}}/{\Omega_{\rm c}}=2,\;{\Omega_{\rm eq}}/{\Omega_{\rm c}}=0.5\right\}$.
		The angular momentum $J$ ranges from 3 to 11 $G M_{\odot}^2 / c$. 
		The $J$-constant turning points are marked with black crosses.
		% All the dynamically evolved models considered here should fall in stable branch, i.e. they all have smaller maximum energy density $\epsilon_{\max}$ than the $J$-constant turning points.
		We find that models with high angular momentum (i.e. $J \gtrsim 9$) fail to preserve their stellar profiles, see figure~\ref{fig:uryu_2.0_0.5_var_e} and \ref{fig:uryu_2.0_0.5_var_j} and the discussions in section~\ref{sec:type_c}.
		\emph{Right panel}: Quasi-spherical (type A) model with $\left\{{\Omega_{\max}}/{\Omega_{\rm c}}=1.6,\;{\Omega_{\rm eq}}/{\Omega_{\rm c}}=1\right\}$.
		The angular momentum $J$ ranges from 3 to 9 $G M_{\odot}^2 / c$. 
		In this type of the models, we do not observe $J$-constant turning points.
		The \texttt{RotNS} code fails to converge when the maximum energy density goes beyond the plotted values, which agrees with~\cite{2022MNRAS.510.2948I}.
		All the dynamically evolved models remain stable up to 20~ms evolutions.
		% Since we did not know if any of these models collapse as black holes, we dynamically evolve models from end to end.
	}
	\label{fig:j_seq}	
\end{figure*}

Figure~\ref{fig:j_seq} shows the gravitational mass $M_{\rm{grav}}$ versus maximum energy density $\epsilon_{\max}$ of constant angular momentum sequences constructed in this work.
Note again that the spatial conformally flat condition is enforced in all the sequences reported here.
% The red and green circles mark the dynamically evolved models, where the red ones collapse as black holes while the green ones do not.
Circles and diamonds mark the dynamically evolved models without introducing perturbations.
The diamonds refer to the models presented in detail in figure~\ref{fig:uryu_2.0_0.5_var_e}, \ref{fig:uryu_2.0_0.5_var_j} and \ref{fig:uryu_1.6_1.0_var_j}.
None of the simulations presented here collapse as black holes.
However, we note that these non-collapsing models are not necessary stable, see further discussions in section~\ref{sec:type_c} and \ref{sec:type_a} below.

The left panel of figure~\ref{fig:j_seq} shows quasi-toroidal (type C) models with $\left\{{\Omega_{\max}}/{\Omega_{\rm c}}=2,\;{\Omega_{\rm eq}}/{\Omega_{\rm c}}=0.5\right\}$, where the angular momentum $J$ ranges from 3 to 11 $G M_{\odot}^2 / c$. 
The $J$-constant turning points are observed in all the quasi-toroidal sequences, which are marked with black crosses in the plot.
According to the turning point criterion~\cite{1988ApJ...325..722F}, the $J$-constant turning points mark the onset of instability.
However, this is not an exact threshold to collapse.
The situation becomes more complicated for differentially rotating cases.
The stability and maximum mass of differentially rotating stars with $J$-constant rotation law~\cite{1988ApJ...325..722F} has been studied~\cite{2018MNRAS.473L.126W, 2019PhRvD.100d3014E, 2023arXiv230206007S}.
More recently, Muhammed~\emph{et~al.}~\cite{2024arXiv240305642M} shows that the turning point criterion seems to also hold in the cases of {Ury{\={u}}} \emph{et~al.}~\cite{2019PhRvD.100l3019U} rotation law.
In this work, we only dynamically evolve the models that have smaller maximum energy density $\epsilon_{\max}$ than the $J$-constant turning points.

The right panel of figure~\ref{fig:j_seq} on the other hand shows the quasi-spherical (type A) model with $\left\{{\Omega_{\max}}/{\Omega_{\rm c}}=1.6,\;{\Omega_{\rm eq}}/{\Omega_{\rm c}}=1\right\}$, where the angular momentum $J$ ranges from 3 to 9 $G M_{\odot}^2 / c$. 
Unlike the quasi-toroidal cases, we do not observe $J$-constant turning points in these sequences.
As the maximum energy density $\epsilon_{\max}$ goes beyond the plotted values, the \texttt{RotNS} code fails to converge.
This behaviour agrees with the discussion in the section~3.2 in~\cite{2022MNRAS.510.2948I}.
The origin of this issue is still unknown, which is beyond the scope of this work and will be investigated in a future study.

\subsection{\label{sec:type_c}Evolutions of quasi-toroidal profiles}
In this subsection, we present some evolutions of the quasi-toroidal (type C) models with different angular momentum $J$ and maximum energy density $\epsilon_{\max}$. 
All the models considered here have lower maximum energy density $\epsilon_{\max}$ than the $J$-constant turning points.
We do not introduce any perturbations into the evolutions.
Since all the low angular momentum cases are found to be stable and trivial, in this section we discuss only the high angular momentum cases (i.e. $J \geq 9$).
% Although all the models considered here are expected to be dynamically stable, the high angular momentum cases are not, which will be discussed in the following.

\begin{figure*}
	\centering
	\includegraphics[width=\textwidth, angle=0]{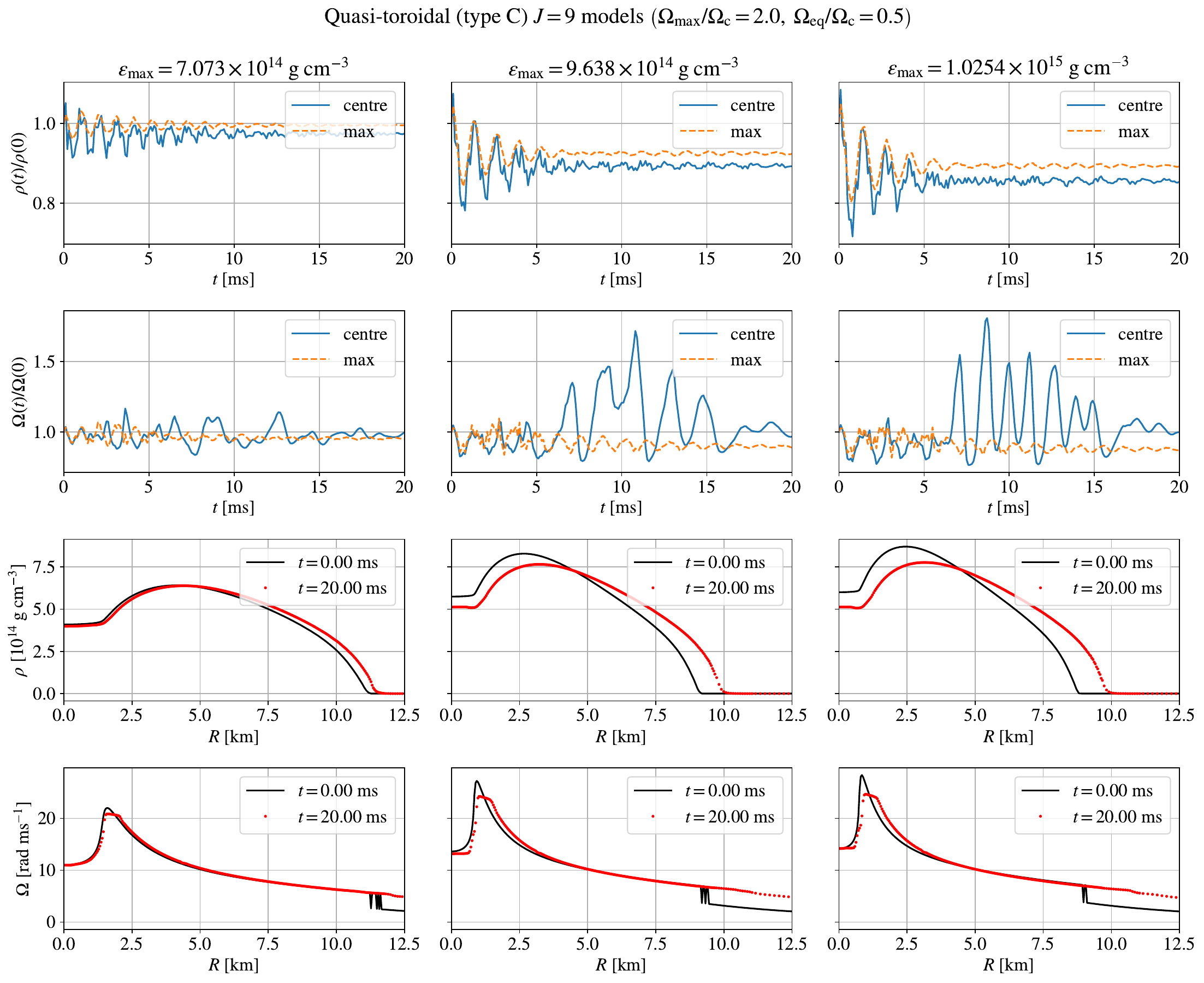}
	\caption{
		Comparison of the dynamical evolutions of the quasi-toroidal (type C) models with angular momentum $J=9$ with different maximum energy densities $\epsilon_{\max}$.
		Three cases are shown in this plot, namely, $\epsilon_{\max} = 7.073\times10^{14}~\rm{g \cdot cm^{-3}}$ (\emph{left column}), $\epsilon_{\max} = 9.638\times10^{15}~\rm{g \cdot cm^{-3}}$ (\emph{middle column}), and $\epsilon_{\max} = 1.0254\times10^{15}~\rm{g \cdot cm^{-3}}$ (\emph{right column}), respectively.
		The \emph{first} and \emph{second rows} show the relative variation of the rest mass densities and angular velocities in time (blue solid lines are for central values while the orange dashed lines are for maximum values).
		In all cases, the rest mass densities oscillate, and gradually settle to a slightly lower value.
		The central angular velocity $\Omega_{\rm c}$ oscillate strongly at around 10 ms, but relax back to initial values later.
		For the high maximum energy density $\epsilon_{\max}$ case, the final central rest mass density $\rho_{\rm{c}}\left(t=20~\rm{ms}\right)$ is about 18\% smaller than the initial value.
		The \emph{third} and \emph{fourth rows} compare of the initial (black solid lines) and final ($t=20 ~\rm{ms}$, red dots) profiles of the rest mass density $\rho$ and the angular velocity $\Omega$ along $R$-axis (i.e. $z=0$).
		The higher the maximum energy density $\epsilon_{\max}$ is, the lower of the final maximum rest mass density $\rho_{\max}$, and stronger the distortion of the rest mass density $\rho$ and the angular velocity $\Omega$ profiles.
		Despite the significant distortions of the rest mass density profiles $\rho\left(R,z=0\right)$, the angular velocity profiles $\Omega\left(R,z=0\right)$ in all the cases considered here are qualitatively preserved.
		}
	\label{fig:uryu_2.0_0.5_var_e}	
\end{figure*}

Figure~\ref{fig:uryu_2.0_0.5_var_e} compares the dynamical evolutions of the quasi-toroidal (type C) models with angular momentum $J=9$ with different maximum energy densities $\epsilon_{\max}$.
% These models have the lowest angular momentum $J$ among all the initial models we have constructed.
In all cases, the maximum rest mass densities $\rho_{\max}$ oscillate, and gradually relax to a slightly lower value.
For the high maximum energy density $\epsilon_{\max} = 1.0245\times10^{15}~\rm{g \cdot cm^{-3}}$ case, the final central rest mass density $\rho_{\rm{c}}\left(t=20~\rm{ms}\right)$ is about 18\% smaller than the initial value, which is not ignorable and indicates that the star migrates into another configuration.
The rest mass density $\rho$ and the angular velocity $\Omega$ along $R$-axis (i.e. $z=0$) at the end of the simulation ($t=20~\rm{ms}$) are significantly distorted except the low maximum energy density case.
The distortions are stronger in the higher maximum energy density $\epsilon_{\max}$ cases.
Although none of these neutron stars collapse to black holes, the medium and high energy density cases were not stable against the evolution up to 20 ms.

\begin{figure*}
	\centering
	\includegraphics[width=\textwidth, angle=0]{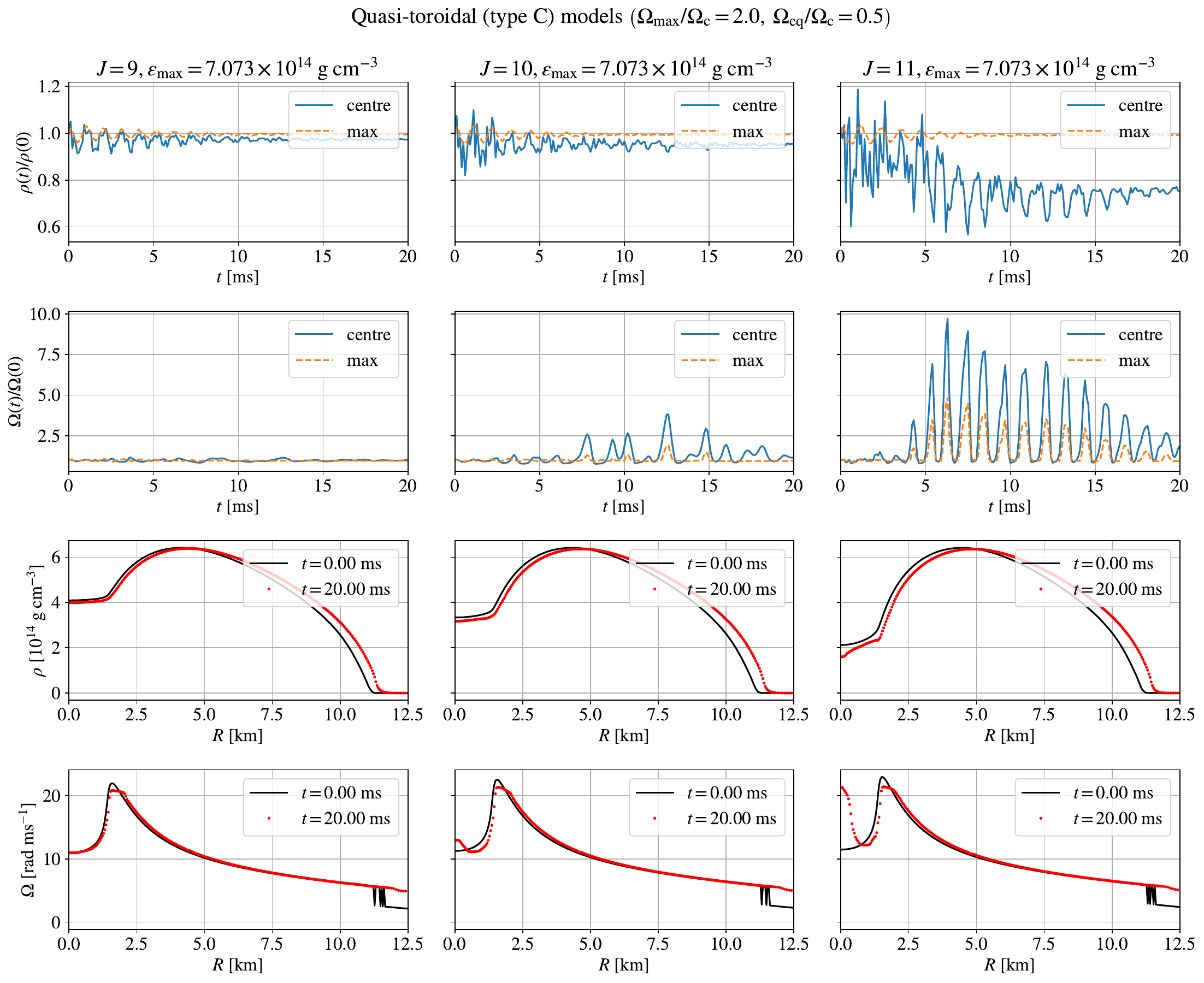}
	\caption{
		Comparison of the dynamical evolutions of the quasi-toroidal (type C) models with maximum energy density $\epsilon_{\max} = 7.073\times10^{14}~\rm{g \cdot cm^{-3}}$ with high angular momentum $J \geq 9$.
		Three cases are shown in this plot, namely, $J=9$ (\emph{left column}), $J=10$ (\emph{middle column}), and $J=11$ (\emph{right column}), respectively.
		The \emph{first} and \emph{second rows} show the relative variation of the rest mass densities and angular velocities in time (blue solid lines are for central values while the orange dashed lines are for maximum values).
		% The \emph{first rows} show the relative variation of the rest mass densities in time (blue solid lines are for central values while the orange dashed lines are for maximum values).
		% The \emph{second rows} show the relative variation of the angular velocities in time.
		The \emph{third} and \emph{fourth rows} compare of the initial (black solid lines) and final ($t=20 ~\rm{ms}$, red dots) profiles of the rest mass density $\rho$ and the angular velocity $\Omega$ along $R$-axis.
		The angular velocity profile $\Omega\left(R,z=0\right)$ for $J > 9$ cases change significantly at the centre of the stars, the rotation law of {Ury{\={u}}} \emph{et~al.}~\cite{2019PhRvD.100l3019U} no longer hold. 
		Moreover, in the highest angular momentum $J=11$ case, although the change of the maximum rest mass density $\rho_{\max}$ is small, the central rest mass density $\rho_{\rm c}$ decrease by about 20\% compare to the initial value.
		Both the central and maximum angular velocities oscillate quasi-periodically.
		The amplitudes of the oscillation are very large, the angular velocities can be a few times higher during the evolutions.
		Note that all the models presented in this plot are the least massive models at a given angular momentum (i.e. far from the $J$-turning points), such strong oscillations and deformations of the stars are unexpected.
		}
	\label{fig:uryu_2.0_0.5_var_j}	
\end{figure*}

Figure~\ref{fig:uryu_2.0_0.5_var_j} compares the dynamical evolutions of also the quasi-toroidal models with maximum energy density $\epsilon_{\max} = 7.073\times10^{14}~\rm{g \cdot cm^{-3}}$ with high angular momentum $J \geq 9$.
The evolutions of the rest mass densities behave similarly, i.e. they oscillate, and gradually relax to a lower value.
The higher the angular momentum $J$ is, the stronger the distortion of the rest mass density $\rho$ and the angular velocity $\Omega$ profiles.
% Although all the models considered here are expected to be dynamically stable, we found that the angular velocity profile $\Omega\left(R,z\right)$ does not preserve in some high angular momentum cases.
Although the maximum energy density of all the models considered here are noticablely lower than the $J$-constant turning points, we found that the angular velocity profiles $\Omega\left(R,z\right)$ are not preserved in some high angular momentum cases.
For instance, in the highest angular momentum $J=11$ case, the central rest mass density $\rho_{\rm c}$ decrease significantly by about 20\% compared to the initial value.
Also, the angular velocity profile $\Omega\left(R,z=0\right)$ changes significantly at the centre of the stars, the rotation law of {Ury{\={u}}} \emph{et~al.}~\cite{2019PhRvD.100l3019U} is violated. 
Both the central and maximum angular velocities oscillate quasi-periodically with large amplitudes, the angular velocities can sometime be a few times higher than the initial values during the evolutions.
Note again that all the models presented in this plot are the least massive models at a given angular momentum (i.e. far from the $J$-turning points), such strong oscillations and deformations of the stars are unexpected.
% Therefore, we do not consider this model remains stable against the evolution.

\subsubsection{Comparison to fully general relativistic cases}
To better understand the origin of the non-stable behaviour, we further compare different combinations of conformally-flat and fully general relativistic initial profiles and evolutions of selected quasi-toroidal (type C) models.
In this section, we focus on the quasi-toroidal (type C) model with maximum energy density $\epsilon_{\max} = 7.073\times10^{14}~\rm{g \cdot cm^{-3}}$ with angular momentum $J = 11$.
The Spectral Einstein Code (\texttt{SpEC})~\cite{SpEC2024} is used for the fully general relativistic evolutions.
The details of the numerical setup in \texttt{SpEC} can be found at~\cite{2020CQGra..37w5010J, 2024arXiv240305642M}.

In this subsection, we consider four cases, namely, 
(i) conformally-flat initial data with conformally-flat evolution;
(ii) general relativistic initial data with conformally-flat evolution;
(iii) conformally-flat initial data with fully general relativistic evolution; and
(iv) general relativistic initial data with fully general relativistic evolution.
One may argue that it is not necessary to consider case~(ii) since it is more-or-less similar to case~(i).  
After all, non-spherically symmetric full general relativistic initial data is in general not conformally flat, there is no self-consistant way to map such initial data into a conformally-flat evolution code.
To evolve such star, we solve the conformally-flat metric equations with the conserved variables of the fully general relativistic star, as discussed in section~\ref{sec:sim}. 
This is effectively enforcing the initial data to be conformally-flat at the beginning. 
However, this does not guarantee that the profile will still be an equilibrium.
In this work, we nevertheless include this case to discuss the validity of mapping a fully general relativistic profile into a conformally-flat evolution code as in~\cite{2023arXiv231211358H}.

\begin{figure}
	\centering
	\includegraphics[width=\columnwidth, angle=0]{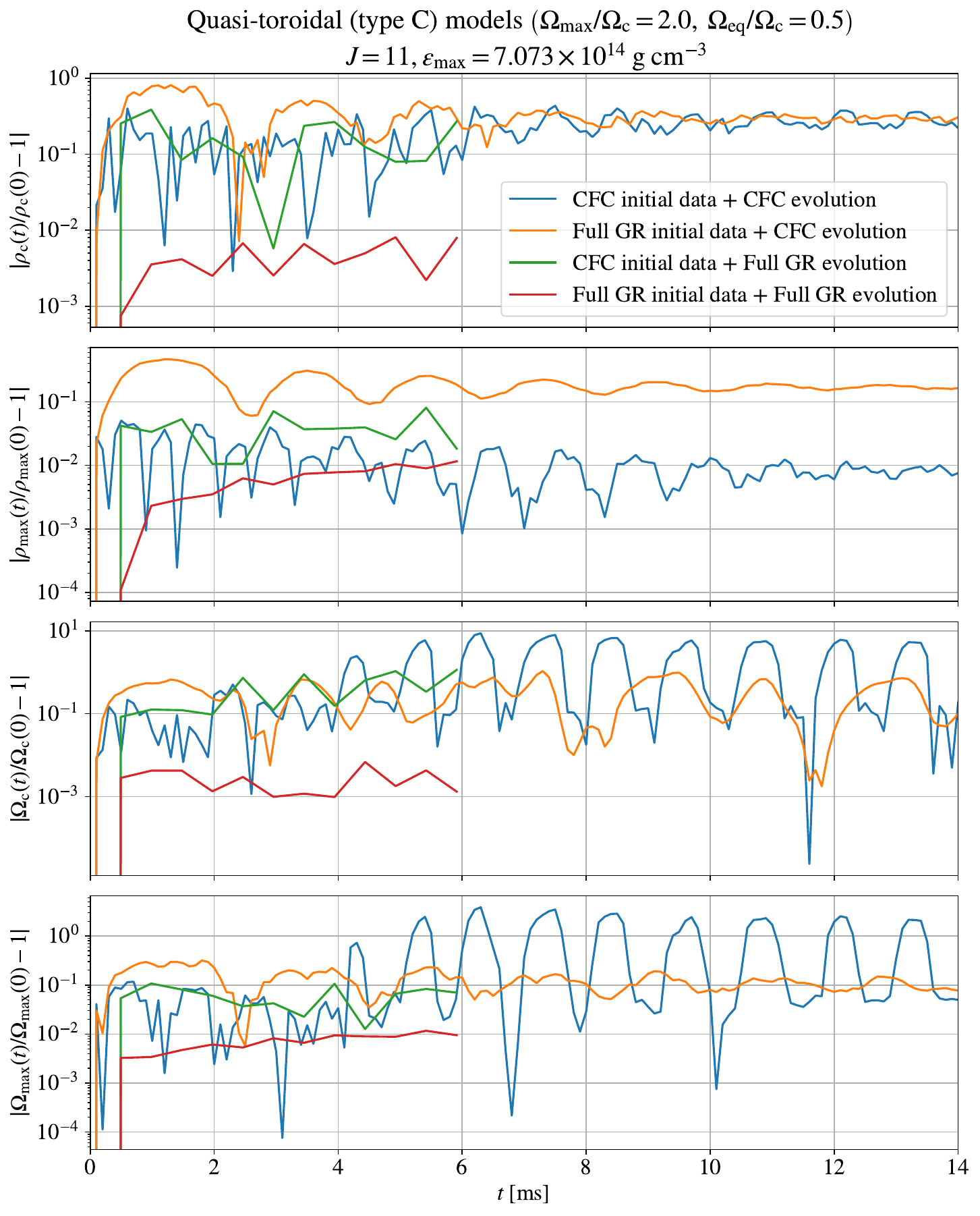}
	\caption{
		The relative variation of the central and maximum rest mass densities and angular velocities of the same models with or without conformally flat approximation.
		The quasi-toroidal (type C) models with maximum energy density $\epsilon_{\max} = 7.073\times10^{14}~\rm{g \cdot cm^{-3}}$ with angular momentum $J = 11$ is used in these simulations.
		All the evolutions have significant deviations except cases where both initial profile and evolution are fully general relativistic (red lines).
		Although the conformally flat equilibrium solutions have only a few percent deviations from their fully general relativistic counterpart~\cite{1996PhRvD..53.5533C, 2014GReGr..46.1800I, 2021MNRAS.503..850I, 2022MNRAS.510.2948I}, the dynamical stabilities of high angular momentum equilibrium models could be changed under conformally flat approximation.
		}
	\label{fig:uryu_2.0_0.5_fullGR}
\end{figure}

\begin{figure}
	\centering
	\includegraphics[width=\columnwidth, angle=0]{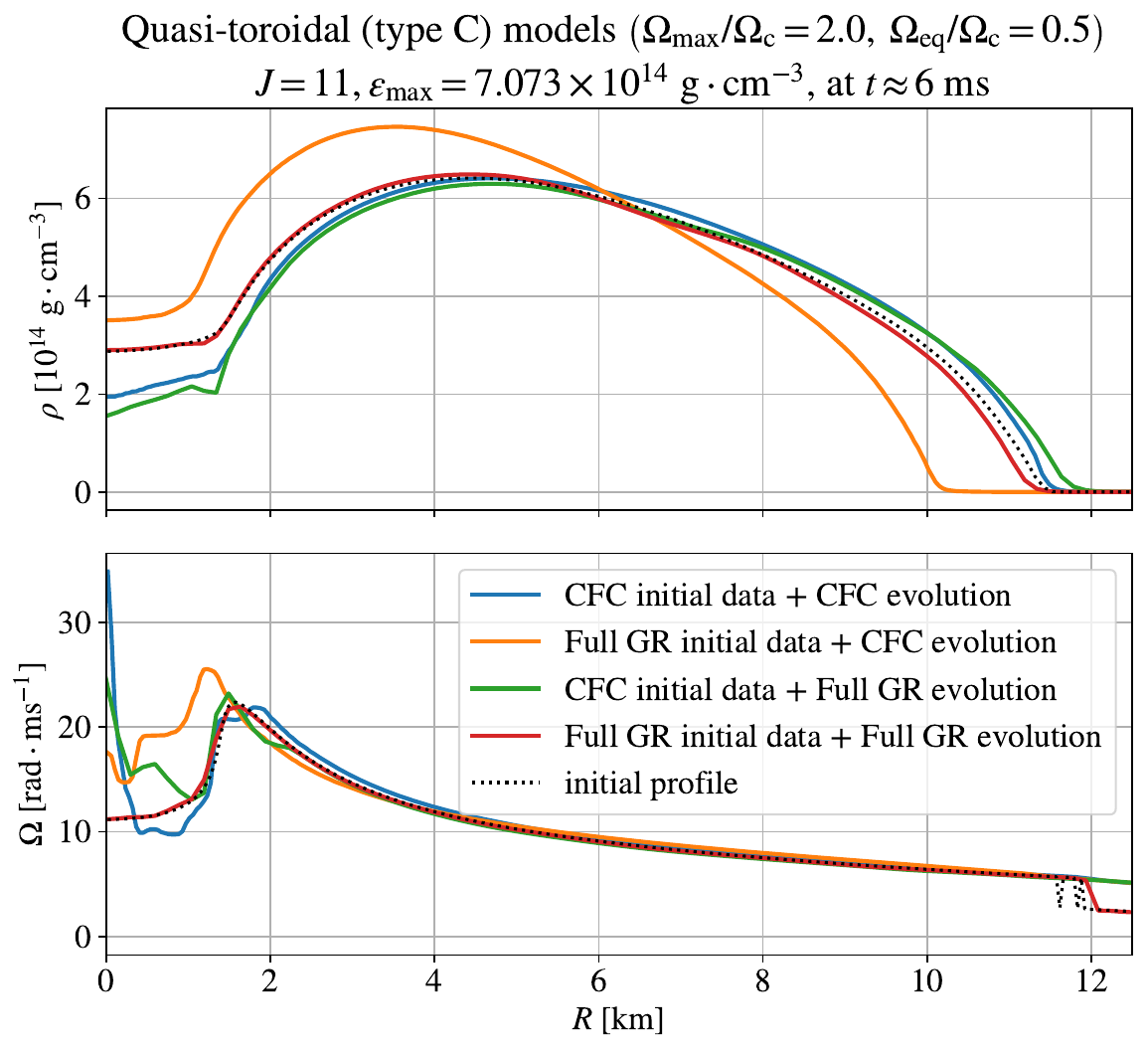}
	\caption{
		Comparison of the rest mass density and angular velocity profiles at the beginning ($t=0~\rm{ms}$, black dotted lines) and at $t\approx 6~\rm{ms}$ (solid lines) of the same models with or without conformally flat approximation.
		The quasi-toroidal (type C) models with maximum energy density $\epsilon_{\max} = 7.073\times10^{14}~\rm{g \cdot cm^{-3}}$ with angular momentum $J = 11$ is used in these simulations.
		The profiles are well-preserved only in the case where both initial profile and evolution are fully general relativistic (red lines).
		}
	\label{fig:uryu_2.0_0.5_full_GR_profile}
\end{figure}

Comparison of the rest mass densities and angular velocities of the same models with or without conformally flat approximation are shown in figure~\ref{fig:uryu_2.0_0.5_fullGR} and \ref{fig:uryu_2.0_0.5_full_GR_profile}.
In particular, figure~\ref{fig:uryu_2.0_0.5_fullGR} shows the absolute values of the relative variation of the central and maximum rest mass densities and angular velocities while figure~\ref{fig:uryu_2.0_0.5_full_GR_profile} compares the profiles at $t\approx 6~\rm{ms}$.
Interestingly, the conformally-flat initial profile is always unstable even with fully general relativistic evolution.
The star remains stable only when the profile and evolution are fully general relativistic (case~(iv), red lines).
This implies that the conformally-flat approximation either makes such high angular momentum star not an equilibrium or makes it an unstable equilibrium.

\subsection{\label{sec:type_a}Evolutions of quasi-spherical profiles}
In this subsection, we present some evolutions of the quasi-spherical (type A) models with different angular momentum $J$ and maximum energy density $\epsilon_{\max}$. 
As mentioned, we do not observe any $J$-constant turning points when we construct the fix angular momentum sequences.
Therefore, we simulate models at both ends, i.e. from low to high maximum energy density $\epsilon_{\max}$.

\begin{figure*}
	\centering
	\includegraphics[width=\textwidth, angle=0]{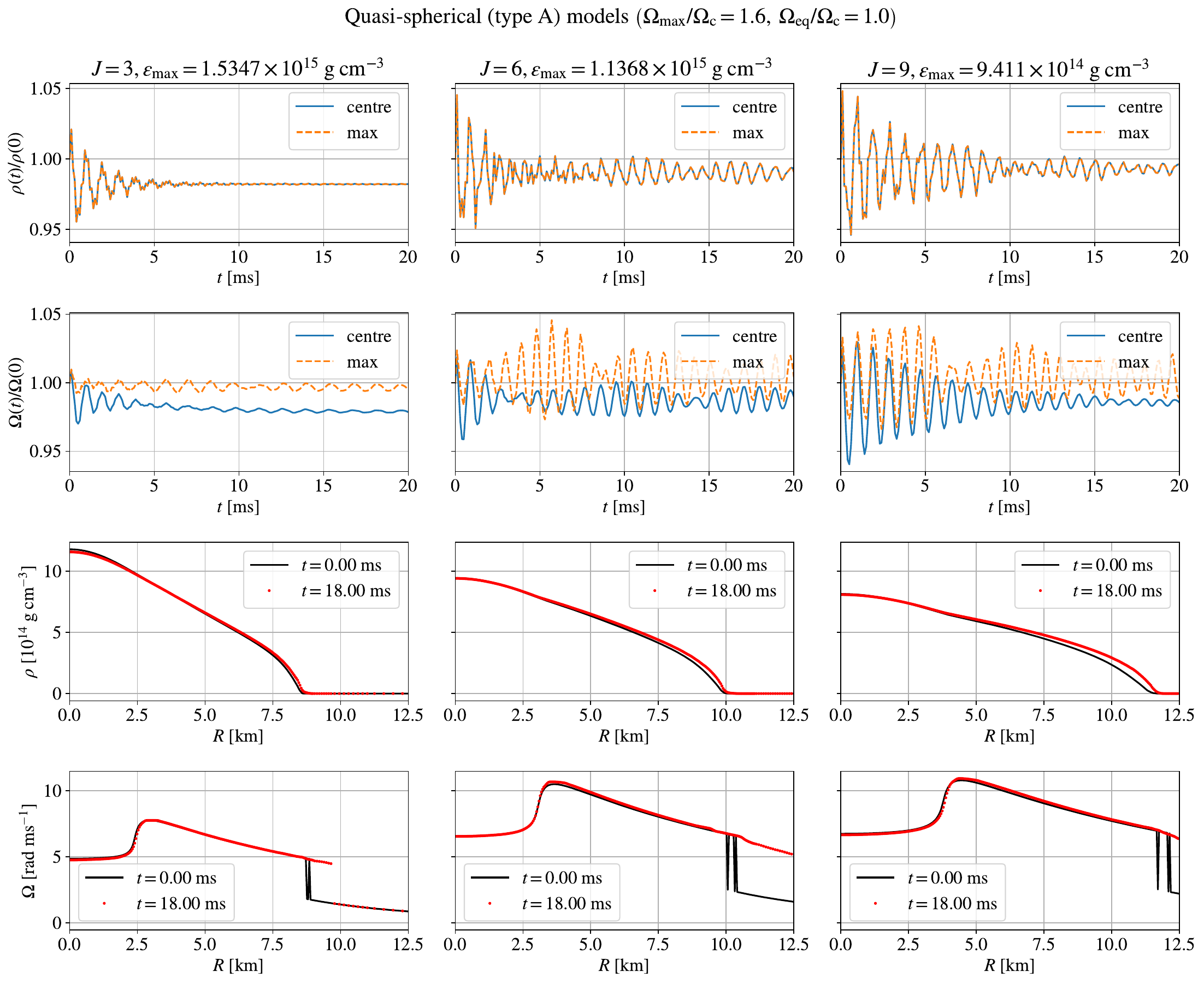}
	\caption{
		Comparison of the dynamical evolutions of the quasi-spherical (type A) models with different maximum energy densities $\epsilon_{\max}$ and angular momentum $J$.
		In this plot, we show the most massive models with three given angular momentum, i.e. $J=3$ (\emph{left column}), $J=6$ (\emph{middle column}), and $J=9$ (\emph{right column}).
		% The \emph{upper panels} show the relative variation of the maximum rest mass density $\rho_{\max}$ in time (blue solid lines), while the \emph{middle} and \emph{lower panel} compare of the initial (black solid lines) and final ($t=10 \;\rm{ms}$, red dots) profiles of the rest mass density $\rho$ and the angular velocity $\Omega$ along $R$-axis.
		The \emph{first} and \emph{second rows} show the relative variation of the rest mass densities and angular velocities in time (blue solid lines are for central values while the orange dashed lines are for maximum values).
		The evolutions of the maximum and central rest mass densities $\rho$ are identical in the quasi-spherical models.
		All the relative variations shown here are within 5\%.
		The \emph{third} and \emph{fourth rows} compare of the rest mass density $\rho$ and the angular velocity $\Omega$ profiles along $R$-axis at the beginning ($t=0~\rm{ms}$, black solid lines) and a later time $t=18~\rm{ms}$ (red dots).
		As shown at the first two rows, the $J=6$ and $J=9$ models are not yet relax to stationary states by the end of the simulations ($t=20 ~ \rm{ms}$), the oscillation is still noticeable in the scale we are plotting.
		Such oscillations result in small quasi-periodic distortions of the stellar profiles (e.g. angular velocity $\Omega$).
		Therefore, instead of plotting the profiles at the end of the simulations, we pick the slightly earlier time (i.e. at $t=18~\rm{ms}$) for better visualisations.
		Unlike the cases in quasi-toroidal (see figure~\ref{fig:uryu_2.0_0.5_var_j} and \ref{fig:uryu_2.0_0.5_var_e}), the profiles are preserved better in this case, and the decrease of the maximum rest mass density $\rho$ is about 2\% even for most massive case with the highest angular momentum $J=9$.
		These results suggest that these quasi-spherical (type A) models are dynamically stable in conformally flat simulations within 20~ms.
		}
	\label{fig:uryu_1.6_1.0_var_j}	
\end{figure*}

Figure~\ref{fig:uryu_1.6_1.0_var_j} compares the dynamical evolutions of the quasi-spherical (type A) models with the most massive models at three given angular momentum $J=3$, $J=6$ and $J=9$.
Although they are the most extreme type A models we have constructed, all of them are dynamically stable in conformally flat simulations.
Indeed, all the simulations we have done (the green circles in the right panel of figure~\ref{fig:j_seq}) of this type of models remain stable.

\section{\label{sec:discussion}Discussion}
% The goal of this work is to investigate the possibilities of using axisymmetric differentially rotating quasi-equilibrium models with high angular momentum to be sensible initial conditions for long term spatially-conformally-flat binary neutron star post-merger modeling.
The goal of this work is to investigate how well the differentially rotating quasi-equilibrium models with high angular momentum remain stable in spatially-conformally-flat simulations.
To this end, we have constructed both quasi-toroidal and quasi-spherical types of spatially-conformally-flat merger-like hot hypermassive neutron stars.
In particular, the ``post-merger-like'' rotation law of {Ury{\={u}}} \emph{et~al.}~\cite{2019PhRvD.100l3019U} is introduced, and assuming constant entropy per baryon $s = 1 ~ k_{\rm{B}} / \text{baryon}$ and in neutrinoless $\beta$-equilibrium.
% In particular, by specifying the ratio between the central, equatorial and maximum angular velocities ($\Omega_{\rm c}$, $\Omega_{\rm eq}$ and $\Omega_{\max}$), we have generated several fix angular momentum sequences of stars in both quasi-toroidal (type C) and quasi-spherical (type A) cases.
We further assess their stability by performing dynamical simulations in conformally flat spacetime using \texttt{Gmunu}.

We show that conformally-flat approximation could alter the dynamical stability of the quasi-toroidal models despite only a few percent difference with their fully general-relativistic variation~\cite{1996PhRvD..53.5533C, 2014GReGr..46.1800I, 2021MNRAS.503..850I, 2022MNRAS.510.2948I}.
Our simulations show that not all conformally-flat quasi-toroidal models remain dynamically stable even for cases where the maximum energy density $\epsilon_{\max}$ is considerably smaller than the $J$-constant turning points.
In high angular momentum (i.e. $J \gtrsim 9$) conformally-flat cases, both the rest mass density and angular velocity can be distorted significantly even with fully general relativistic evolutions.
However, this is not the case when both initial profile and evolutions are fully general-relativistic.
This implies that conformally-flat approximation either makes such high angular {momentum} star not an equilibrium or makes it an unstable equilibrium.
% We found that conformally-flat approximation could alter the stabilities of the high angular momentum quasi-toroidal stars, and hence has the limitation.
Mapping these stellar profiles from fully general relativistic simulations to other codes by assuming conformally-flat conditions (e.g.~\cite{2017ApJ...846..114F, 2023arXiv231211358H}) could result in very different lifetime of the star, and therefore affecting the modeling of the matter outflow.
{
The origin of such behaviour can be better understood by studying its hydrodynamical instability (e.g. \cite{2018MNRAS.475L.125G}), which is left as future work.
}

On the other hand, unlike the quasi-toroidal models, we show that all the quasi-spherical models considered in this work remain stable.
% Indeed, whether a differentially rotating star is dynamically unstable is non-trivial, it relates to many factors, including rotation laws and types of solutions~\cite{2018MNRAS.473L.126W, 2019PhRvD..99h3017E, 2020PhRvD.102d4040X, 2023arXiv230206007S}.
% Nevertheless, this will not prevent us to model long-lived hypermassive neutron stars by using quasi-equilibrium models under the conformal flatness approximation.
The quasi-spherical models are not only by construction more post-merger-like compared to the quasi-toroidal models, they are dynamically stable even for the most extreme cases we considered.
These properties make them ideal choices for long-lived hypermassive neutron star modeling.
% We show that the quasi-spherical models are more suitable to be used as initial data for long-lived hypermassive neutron star modeling in conformally flat spacetime.
Generating different sequences with different parameters (e.g. mass, angular momentum, equation of state) enable us to systemically study how these parameters affect the outcomes of the post-merger.
In the future, we will attempt to deliver post-merger modeling with such quasi-spherical models together with magnetic fields and neutrino transport.

\begin{acknowledgments}
P.C.K.C. gratefully acknowledges support from NSF Grant PHY-2020275 (Network for Neutrinos, Nuclear Astrophysics, and Symmetries (N3AS)).
F.F. gratefully acknowledge support from the Department of Energy, Office of Science, Office of Nuclear Physics, under contract number DE-AC02-05CH11231 and from the NSF through grant AST-2107932. 
M.D. gratefully acknowledges support from the NSF through grant PHY-2110287.  
M.D. and F.F. gratefully acknowledge support from NASA through grant 80NSSC22K0719. 

The simulations in this work have been performed on the third UNH supercomputer Marvin, also known as Plasma, which is supported by NSF/MRI program under grant number AGS-1919310. 
The data of the simulations were post-processed and visualised with 
\texttt{yt}~\citep{2011ApJS..192....9T},
\texttt{NumPy}~\citep{harris2020array}, 
\texttt{pandas}~\citep{reback2020pandas, mckinney-proc-scipy-2010},
\texttt{SciPy}~\citep{2020SciPy-NMeth} and
\texttt{Matplotlib}~\citep{2007CSE.....9...90H, thomas_a_caswell_2023_7697899}.
\end{acknowledgments}

\appendix
\section{\label{sec:convergence}Convergence tests}
\begin{figure}
	\centering
	\includegraphics[width=\columnwidth, angle=0]{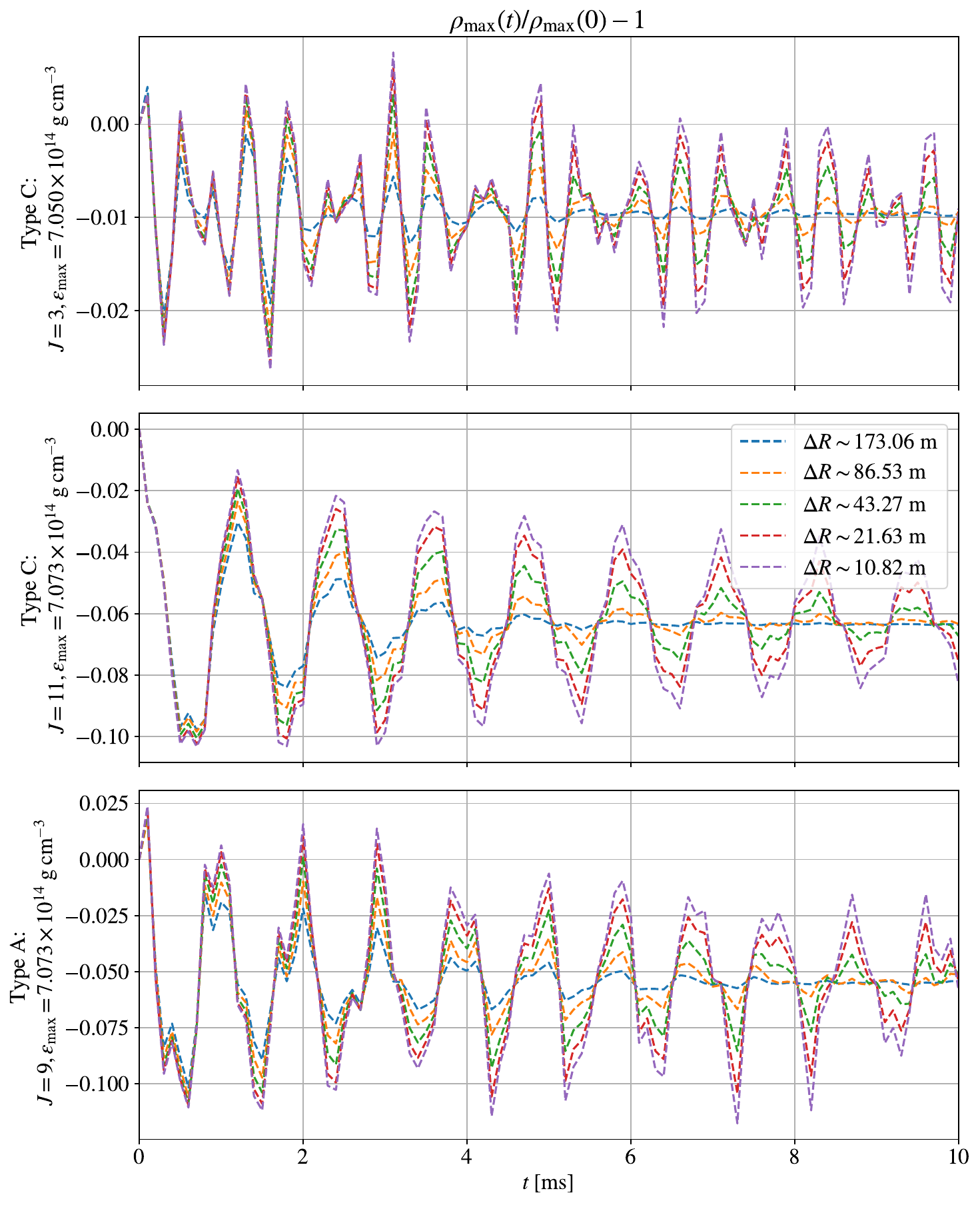}
	\caption{
		The relative variation of the maximum rest mass density $\rho_{\max}$ with different resolution in different models.
		\emph{Upper panel}: Quasi-toroidal (type C) model with $J=3~\epsilon_{\max} = 7.050\times10^{14}~\rm{g \cdot cm^{-3}}$.
		\emph{Middle panel}: Quasi-toroidal (type C) model with $J=11~\epsilon_{\max} = 7.073\times10^{14}~\rm{g \cdot cm^{-3}}$.
		\emph{Lower panel}: Quasi-spherical (type A) model with $J=9~\epsilon_{\max} = 7.073\times10^{14}~\rm{g \cdot cm^{-3}}$.
		Initial in-going velocity perturbation is introduced in these simulations.
		As shown in the plot, the oscillation amplitudes are mostly the same at the very beginning, and gradually decrease as time goes on.
		At low resolution, the simulations are very diffusive, which relax the oscillation sooner.
		}
	\label{fig:conv_tests}	
\end{figure}

Here we present the convergence tests.
The simulations have the same setup as in the paper, except that we introduce initial in-going velocity perturbation.
As shown in figure~\ref{fig:conv_tests}.
As shown in the plot, the oscillation amplitudes are mostly the same at the very beginning, and gradually decrease as time goes on.
At low resolution, the simulations are very diffusive, which relax to the stationary state sooner.
%fixme: The crucial question for distinguishing non-equilibrium from unstable equilibrium is whether, without perturbation, the initial time derivative converge to zero with resolution (as the truncation error perturbation is reduced).  For that reason, it would be useful to zoom in on the very early evolution.  Is there anything--fluid or metric variable--with a converged non-zero initial time derivative?

%fixme: From this plot, it looks like all of these have some sort of perturbation in them.  None of them (A or C type) are exact equilibria.  Then the question is, how can this be?  Hydrostatic equilibrium for the fluid variables is imposed by the Bernouilli condition, if you zoom in enough, you should see the density time derivative go to zero, up to issues with interpolating from a finite-resolution RotNS profile maybe.  I could believe that the metric time derivatives could be initially non-zero because of your conformal flatness alterations, but you haven't isolated this.

% \nocite{*}

% \bibliographystyle{apsrev4-2}
% \bibliographystyle{aas_marcos}
\bibliography{references}{}% Produces the bibliography via BibTeX.

\end{document}